\newcommand{\CLASSINPUTtoptextmargin}{0.75in}%
\newcommand{\CLASSINPUTbottomtextmargin}{1in}%
\newcommand{\CLASSINPUTinnersidemargin}{0.63in}%
\newcommand{\CLASSINPUToutersidemargin}{0.63in}%
\documentclass[conference,10pt,letterpaper]{IEEEtran}%
\usepackage{amsmath}
\usepackage{amssymb}
\usepackage{graphicx}
\graphicspath{{figures/}}
\usepackage{multirow}
\usepackage[none]{hyphenat}
\usepackage{float}
\usepackage{subfig}
\usepackage{dblfloatfix}
\usepackage{t1enc}
\usepackage{times}
\usepackage{acronym}
\usepackage{siunitx}
\DeclareSIUnit{\sample}{Sa}
\DeclareMathOperator*{\argmax}{max}
\makeatletter

\def\@IMSauthorblockNAMEstyle{\normalfont\IMSauthorsize}
\def\@IMSauthorblockAFFILstyle{\normalfont\IMSaffilsize}
\def\@IMSauthorblockEMAILstyle{\normalfont\IMSaffilsize}
\def\IMSauthorblockNAME#1{%
\relax\@IMSauthorblockNAMEstyle%
#1%
}%
\def\IMSauthorblockAFFIL#1{%
\relax\@IMSauthorblockAFFILstyle%
\vskip\@IEEEauthorblockAtopspace
#1%
}%
\def\IMSauthorblockEMAIL#1{%
\relax\@IMSauthorblockEMAILstyle%
\vskip\@IEEEauthorblockAtopspace
#1%
}%
\newcommand{\IMSauthor}[1]{%
\ifIsBlindReviewVersion%
\author{\phantom{\parbox{\textwidth}{\center\relax#1}}}%
\else%
\author{\parbox{\textwidth}{\center\relax#1}}%
\fi%
}%
\newif\ifIsBlindReviewVersion
\def\IMSthispaperforblindreview{\IsBlindReviewVersiontrue}
\def\IMSthispaperforfinalpublication{\IsBlindReviewVersionfalse}
\def\@maketitle{\newpage
\bgroup\par\addvspace{0.5\baselineskip}\centering%
\ifCLASSOPTIONtechnote
   {\bfseries\large\@IEEEcompsoconly{\sffamily}\@title\par}\vskip 1.3em{\lineskip .5em\@IEEEcompsoconly{\sffamily}\@author
   \@IEEEspecialpapernotice\par{\@IEEEcompsoconly{\vskip 1.5em\relax
   \@IEEEtitleabstractindextextbox{\@IEEEtitleabstractindextext}\par
   \hfill\@IEEEcompsocdiamondline\hfill\hbox{}\par}}}\relax
\else
   \vskip0.2em{\IMStitlesize\ifCLASSOPTIONtransmag\bfseries\LARGE\fi\@IEEEcompsoconly{\sffamily}\@IEEEcompsocconfonly{\normalfont\normalsize\vskip 2\@IEEEnormalsizeunitybaselineskip
   \bfseries\Large}\@title\par}\vskip1.0em\par
   \ifCLASSOPTIONconference%
      {\@IEEEspecialpapernotice\mbox{}\vskip\@IEEEauthorblockconfadjspace%
       \mbox{}\hfill\begin{@IEEEauthorhalign}\@author\end{@IEEEauthorhalign}\hfill\mbox{}\par}\relax
   \else
      \ifCLASSOPTIONpeerreviewca
         {\@IEEEcompsoconly{\sffamily}\@IEEEspecialpapernotice\mbox{}\vskip\@IEEEauthorblockconfadjspace%
          \mbox{}\hfill\begin{@IEEEauthorhalign}\@author\end{@IEEEauthorhalign}\hfill\mbox{}\par
          {\@IEEEcompsoconly{\vskip 1.5em\relax
           \@IEEEtitleabstractindextextbox{\@IEEEtitleabstractindextext}\par\hfill
           \@IEEEcompsocdiamondline\hfill\hbox{}\par}}}\relax
      \else
         \ifCLASSOPTIONtransmag
           {\@IEEEspecialpapernotice\mbox{}\vskip\@IEEEauthorblockconfadjspace%
            \mbox{}\hfill\begin{@IEEEauthorhalign}\@author\end{@IEEEauthorhalign}\hfill\mbox{}\par
           {\vspace{0.5\baselineskip}\relax\@IEEEtitleabstractindextextbox{\@IEEEtitleabstractindextext}\vspace{-1\baselineskip}\par}}\relax
         \else
           {\lineskip.5em\@IEEEcompsoconly{\sffamily}\sublargesize\@author\@IEEEspecialpapernotice\par
           {\@IEEEcompsoconly{\vskip 1.5em\relax
            \@IEEEtitleabstractindextextbox{\@IEEEtitleabstractindextext}\par\hfill
            \@IEEEcompsocdiamondline\hfill\hbox{}\par}}}\relax
         \fi
      \fi
   \fi
\fi\par\addvspace{0.0\baselineskip}\egroup}

\def\IMStitlesize{\@setfontsize{\IMStitlesize}{18}{21pt}}
\def\IMSauthorsize{\@setfontsize{\IMSauthorsize}{12}{13pt}}
\def\IMSaffilsize{\@setfontsize{\IMSaffilsize}{12}{13pt}}
\def\IMScaptionsize{\@setfontsize{\IMScaptionsize}{8}{9pt}}
\def\IMSbibsize{\@setfontsize{\IMSbibsize}{8}{9pt}}

\def\@IEEEauthorblockNstyle{\IMSauthorsize\@IEEEcompsocnotconfonly{\sffamily}\@IEEEcompsocconfonly{\large}}
\def\@IEEEauthorblockAstyle{\IMSaffilsize\@IEEEcompsocnotconfonly{\sffamily}\@IEEEcompsocconfonly{\itshape}\@IEEEcompsocconfonly{\large}}
\def\@IEEEauthordefaulttextstyle{\IMSauthorsize\@IEEEcompsocnotconfonly{\sffamily}\sublargesize}

\def\thebibliography#1{\section*{\refname}%
    \addcontentsline{toc}{section}{\refname}%
    \IMSbibsize\@IEEEcompsocconfonly{\small}\vskip 0.3\baselineskip plus 0.1\baselineskip minus 0.1\baselineskip
    \list{\@biblabel{\@arabic\c@enumiv}}%
    {\settowidth\labelwidth{\@biblabel{#1}}%
    \leftmargin\labelwidth
    \advance\leftmargin\labelsep\relax
    \itemsep \IEEEbibitemsep\relax
    \usecounter{enumiv}%
    \let\p@enumiv\@empty
    \renewcommand\theenumiv{\@arabic\c@enumiv}}%
    \let\@IEEElatexbibitem\bibitem%
    \def\bibitem{\@IEEEbibitemprefix\@IEEElatexbibitem}%
\def\newblock{\hskip .11em plus .33em minus .07em}%
\ifCLASSOPTIONtechnote\sloppy\clubpenalty4000\widowpenalty4000\interlinepenalty100%
\else\sloppy\clubpenalty4000\widowpenalty4000\interlinepenalty500\fi%
    \sfcode`\.=1000\relax}

\long\def\@makecaption#1#2{%
\ifx\@captype\@IEEEtablestring%
\par\@IEEEtabletopskipstrut
\else
\@IEEEfigurecaptionsepspace
\fi
\setbox\@tempboxa\hbox{\normalfont\IMScaptionsize {#1.}\nobreakspace\nobreakspace #2}%
\ifdim \wd\@tempboxa >\hsize%
\setbox\@tempboxa\hbox{\normalfont\IMScaptionsize {#1.}\nobreakspace\nobreakspace}%
\parbox[t]{\hsize}{\normalfont\IMScaptionsize\noindent\unhbox\@tempboxa#2}%
\else
\ifCLASSOPTIONconference \hbox to\hsize{\normalfont\IMScaptionsize\hfil\box\@tempboxa\hfil}%
\else \hbox to\hsize{\normalfont\IMScaptionsize\box\@tempboxa\hfil}%
\fi\fi
\ifx\@captype\@IEEEtablestring%
\@IEEEtablecaptionsepspace
\else
\fi}

\newlength\tablecaptiontotableskip
\newlength\figuretocaptionskip
\def\@IEEEfigurecaptionsepspace{\vskip\figuretocaptionskip\relax}%
\def\@IEEEtablecaptionsepspace{\vskip\tablecaptiontotableskip\relax}%

\def\abstract{\normalfont%
\@IEEEabskeysecsize\bfseries\textit{\abstractname}\,\bfseries\textit{---}\,%
\@IEEEgobbleleadPARNLSP}%

\def\IEEEkeywords{\normalfont%
\@IEEEabskeysecsize\bfseries\textit{\IEEEkeywordsname}\,\bfseries\textit{---}\,%
\@IEEEgobbleleadPARNLSP}%
\def\endIEEEkeywords{\relax\vspace{0.67ex}%
\par\if@twocolumn\else\endquotation\fi%
\normalsize\normalfont}%

\DeclareRobustCommand*{\IMSauthorrefmark}[1]{\raisebox{0pt}[0pt][0pt]{\textsuperscript{\footnotesize{#1}}}}%
\def\@IEEEauthorblockNtopspace{0ex}
\def\@IEEEauthorblockAtopspace{1mm}
\def\tablename{Table}
\def\thetable{\arabic{table}}
\def\IEEEkeywordsname{Keywords}
\def\subsubsection{\@startsection{subsubsection}{3}{\z@}{1.5ex plus 1.5ex minus 0.5ex}%
{0.7ex plus .5ex minus 0ex}{\normalfont\normalsize\itshape}}%
\def\@seccntformat#1{\csname the#1dis\endcsname\relax}
\def\thesectiondis{\thesection.\hskip 0.5em}
\def\thesubsectiondis{{\hbox to\parindent{\Alph{subsection}.}}}
\def\thesubsubsectiondis{{\hbox to \parindent{\arabic{subsubsection})}}}
\def\theparagraphdis{{\hbox to \parindent{\alph{paragraph})}}}
\IEEEilabelindentA \parindent
\IEEEilabelindent \IEEEilabelindentA
\IEEEelabelindent \parindent
\IEEEdlabelindent \parindent
\IEEElabelindent \parindent

\newlength\@IMSparindent
\newcommand\IMSdisplayacksection[1]{%
\ifIsBlindReviewVersion%
\noindent\phantom{\parbox[t]{\columnwidth}{\normalbaselines\setlength{\parindent}{\@IMSparindent}{#1}\strut}}
\else%
\noindent\parbox[t]{\columnwidth}{\normalbaselines\setlength{\parindent}{\@IMSparindent}{#1}\strut}%
\fi%
}%

\makeatother

\begin{document}
\raggedbottom
%
%
%
\title{Direction Finding for Software Defined Radios with Switched Uniform Circular Arrays}
\IMSthispaperforblindreview
\IMSthispaperforfinalpublication
\IMSauthor{%
\IMSauthorblockNAME{
Lennart Werner\IMSauthorrefmark{\#1},
Markus Gardill\IMSauthorrefmark{\$2},
Marco Hutter\IMSauthorrefmark{\#3}, 
}
\IMSauthorblockAFFIL{
\IMSauthorrefmark{\#}Robotic Systems Lab, ETH Zürich, Switzerland\\
\IMSauthorrefmark{\$}Chair of Electronic Systems and Sensors, BTU Cottbus-Senftenberg, Germany\\
}
\IMSauthorblockEMAIL{
\{\IMSauthorrefmark{1}lwerne, \IMSauthorrefmark{3}mahutter\}@ethz.ch, \IMSauthorrefmark{2}markus.gardill@b-tu.de
}
}
\acrodef{RL}{Reinforcement Learning}
\acrodef{NN}{Neural Network}
\acrodef{DoF}{Degree of Freedom}
\acrodef{ID}{Inverse Dynamics}
\acrodef{IMU}{Inertial Measurement Unit}
\acrodef{PPO}{Proximal Policy Optimization}
\acrodef{PSD}{Power Spectral Density}
\acrodef{MLP}{Multi-Layer Perceptron}
\acrodef{COG}{Center Of Gravity}
\acrodef{EE}{end effector}
\acrodef{DoA}{direction of arrival}
\acrodef{RF}{radio frequency}
\acrodef{SDR}{Software Defined Radio}
\acrodef{SNR}{Signal to Noise Ratio}
\acrodef{SLL}{Sidelobe Level}
\acrodef{SM}{Switching Matrix}
\acrodef{CW}{Continuous Wave}
\acrodef{ISM}{Industrial, Scientific, and Medical}
\acrodef{TDM}{Time Devision Multiplexing}
\acrodef{UCA}{Uniform Circular Array}
\acrodef{ULA}{Uniform Linear Array}
%
\maketitle
\begin{abstract}

Accurate Direction of Arrival (DoA) estimation is critical for applications in robotics and communication, but high costs and complexity of coherent multi-channel receivers hinder accessibility. 
This work proposes a cost-effective DoA estimation system for continuous wave (CW) signals in the 2.4 GHz ISM band. 
A two-channel software-defined radio (SDR) with time-division multiplexing (TDM) enables pseudo-coherent sampling of an eight-element uniform circular array (UCA) with low hardware complexity. 
A central reference antenna mitigates phase jitter and sampling errors.

The system applies an enhanced MUSIC algorithm with spatial smoothing to handle light multipath interference in indoor and outdoor environments. 
Experiments in an anechoic chamber validate accuracy under ideal conditions, while real-world tests confirm robust performance in multipath-prone scenarios. 
With 5 Hz DoA updates and post-processing to enhance tracking, the system provides an accessible and reliable solution for DoA estimation in real-world environments.
\end{abstract}

\section{Introduction}
A wide range of industrial and scientific applications require accurate and reliable \ac{DoA} estimation.
It enables new capabilities such as indoor localization \cite{8905238} and leader-following of quadruped robots in crowded environments \cite{scheidemann2024obstacle}. 
However, the high cost of coherent multi-channel receivers for \ac{DoA} measurements often restricts their broader adoption in research and practical applications \cite{dai2022deepaoanet}.
This leads to a lack of real-world experimental validation for many processing techniques.

\begin{figure}
    \centering
    \includegraphics[width=1\linewidth]{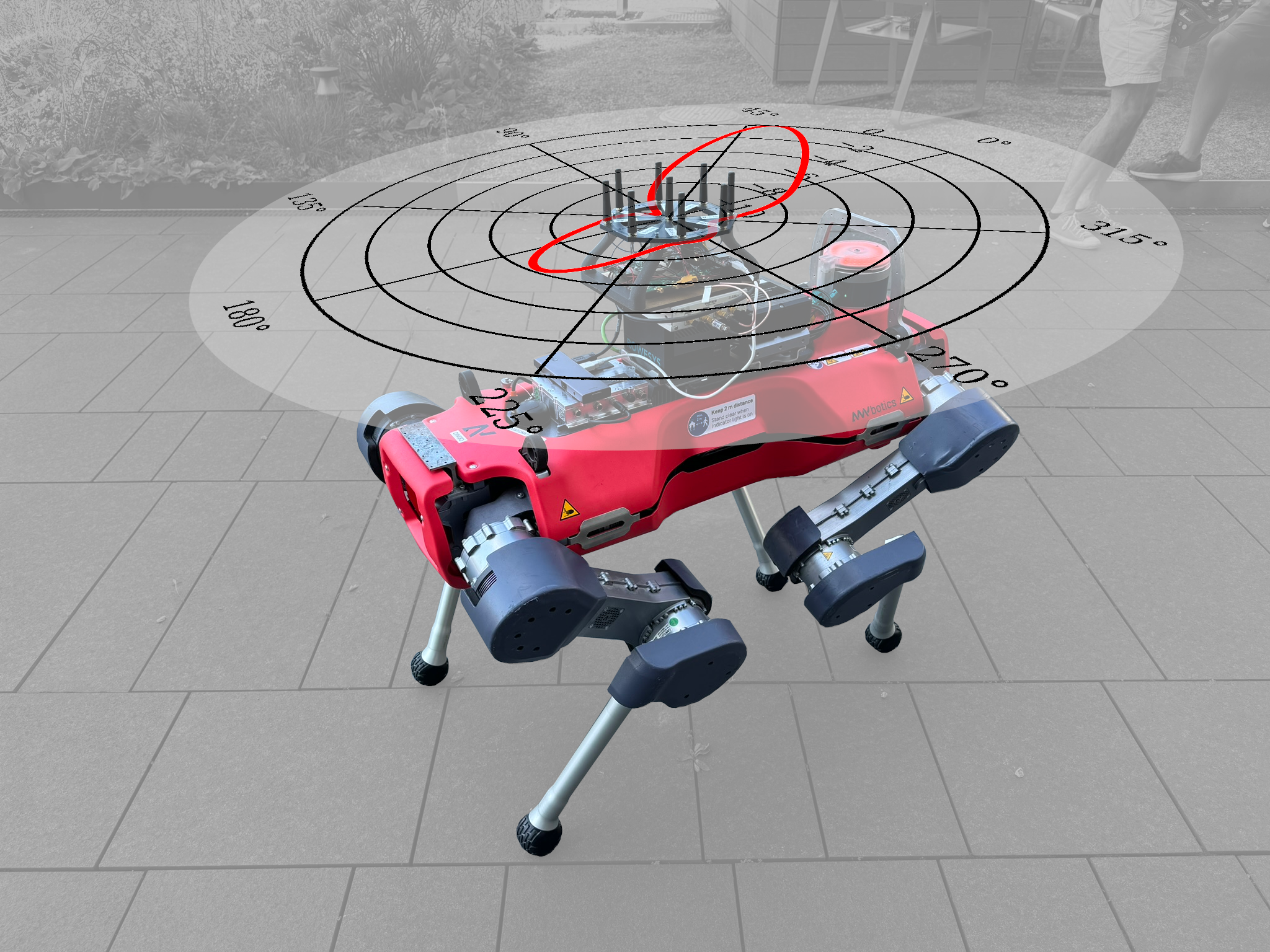}
    \caption{Real world experiment setup. 8+1 \ac{UCA} with switching matrix and \ac{SDR} for \ac{DoA} estimation mounted on a quadrupedal robot operating in dense urban environment.}
    \label{fig:anymal_aoa}
\end{figure}

The main contribution of this paper is the development of a cost-effective and practical hardware and processing pipeline for \ac{DoA} estimation using enhanced MUSIC~\cite{schmidt1986multiple}. 
A dual-channel \ac{SDR} combined with an inexpensive switching matrix enables pseudo-coherent sampling of an $8+1$ channel \ac{UCA}.
We locate the \ac{DoA} of a \ac{CW} transmitter in the \SI{2.4}{\GHz} \ac{ISM} band. 
The system is characterized through anechoic chamber measurements and validated in real-world indoor and outdoor scenarios under light multi-path conditions and mutual coupling (Fig.~\ref{fig:anymal_aoa}).

Building on the Dual-Channel Cyclic MUSIC approach \cite{8695705}, we extend its application to \acp{UCA} and introduce spatial smoothing to manage correlated signals. 
Inspired by \cite{299559}, we compensate for mutual coupling within the array.
MUSIC was selected for its superior stability and accuracy compared to methods like minimum-norm and ESPRIT~\cite{lavate2010performance}. 
By adding spatial smoothing, the algorithm can detect a limited number of coherent signals. 
Its sensitivity to array response errors is addressed by a careful characterization conducted in an anechoic chamber.

Alternative methodologies have successfully integrated MUSIC with rank reduction techniques to mitigate the adverse effects of mutual coupling, while enabling the estimation of the elevation angle~\cite{goossens2007hybrid}. 
Methods not based on subspace approaches, such as SAGE~\cite{fleury1996wideband}, are capable of handling coherent signal sources without necessitating spatial smoothing.
Alternative maximum-likelihood methods exhibit comparable benefits, offering different strategies to manage the heightened computational demands~\cite{stoica1990maximum}. 
Recently, learning-based approaches have demonstrated promising results in terms of resilience to array imperfections~\cite{liu2018direction, fuchs2022machine}, yielding enhanced precision relative to traditional techniques. 
However, the capability of learning methods to generalize effectively in environments characterized by significant multi-path interference remains an unresolved challenge~\cite{pisa2024generalization}.

In this paper, we adopt the following notations: The superscripts $*$, $T$, and $H$ represent the conjugate, transpose, and Hermitian transpose, respectively. 
The symbols $E\{\cdot\}$, $\odot$, $\cdot$, and $||\cdot||$ denote the expectation, Hadamard product, matrix product, and matrix norm. 
$\mathbf{D(x)}$ is used to indicate a diagonal matrix with $\mathbf{x}$ along its main diagonal.
Additionally, $\mathbf{I_M}$ refers to the $M \times M$ identity matrix unless stated otherwise.
$\mathbb{C}$ is the group of complex numbers.
For evaluation, we use the \ac{SLL} as the ratio of the correct signal to the next strongest incorrect peak.

\section{Hardware Setup}
\begin{figure}
    \centering
    \includegraphics[width=0.7\linewidth]{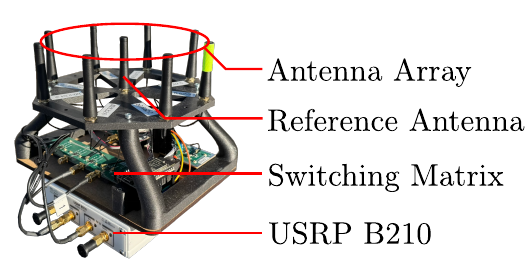}
    \caption{Measurement setup with \ac{UCA}, switching matrix and USRP.}
    \label{fig:setup}
\end{figure}

As a target beacon, we use a simple signal generator transmitting a \ac{CW} tone at $f_{\mathrm{Tx}} = \SI{2.4}{\GHz}$.
The receiving antenna array is an $8+1$ \ac{UCA} with a radius of $0.5\lambda$ with $\lambda = c/f$.
The central reference antenna is directly connected to one port of the USRP B-210~\cite{ettusUSRPB210}.
Port two is connected to the eight circular antennas through a one-to-eight switching matrix as visible in Fig.~\ref{fig:schematic}.
Fig.~\ref{fig:setup} shows the main components.

\subsection{USRP}
\begin{figure}
    \centering
    \scalebox{0.55}{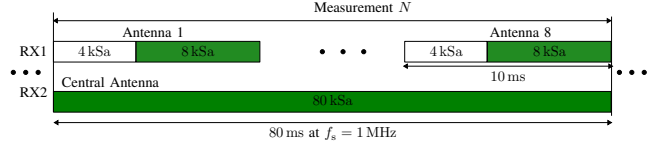}
    \caption{Timing of the data acquisition.}
    \label{fig:timing}
\end{figure}

Timed commands, as well as fixed gain, ensure receiver time-coherence and consistency between measurements.
As illustrated in Fig.~\ref{fig:timing}, each measurement chunk consists of \SI{80}{\kilo\sample} obtained at a sampling rate of  $f_{\mathrm{s}} = \SI{1}{\MHz}$, spanning all switching configurations.
A margin of \SI{4}{\kilo\sample} between each switching event is used to avoid transient effects in the processed data.
The remaining \SI{6}{\kilo\sample} samples per antenna are used for processing with the associated samples from the central antenna as a reference.
The Rx frequency is selected to be $f_{\mathrm{Rx}} = f_{\mathrm{Tx}} - \SI{50}{\kHz}$, such that the received signal is present at an IF frequency of $f_{\mathrm{IF}} = \SI{50}{\kilo\hertz}$ in order to move the signal away from the DC bias.
 
\subsection{Switching Matrix}
\begin{figure}
    \centering
    \includegraphics[width=.7\linewidth]{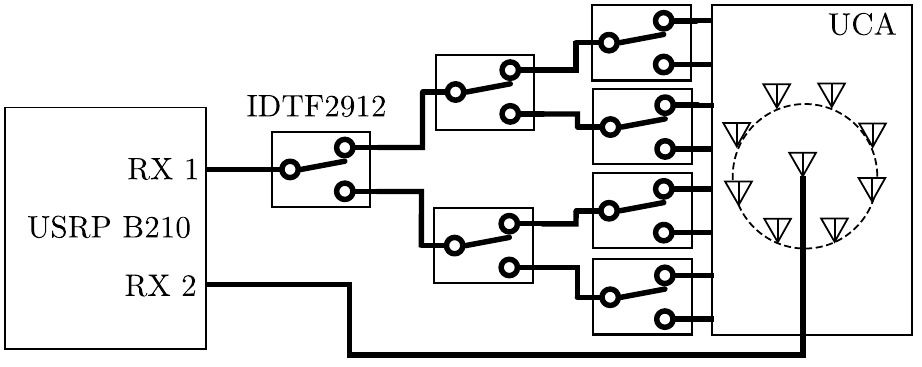}
    \caption{Schematic of the proposed measurement setup.}
    \label{fig:schematic}
\end{figure}

The Switching matrix in Fig. \ref{fig:schematic} is built with seven IDTF2912 \ac{RF} switches.
A microcontroller toggles the \SI{10}{\ms} timed switching sequence after a single USB trigger.
The maximum phase error of the switching matrix is $\pm \SI{3}{\degree}$ at \SI{2.4}{\GHz} between all ports.

\section{Signal Processing}
The processing is split into the stages: Pseudo coherence recovery, array compensation, and MUSIC evaluation.
\subsection{Coherence Recovery}
Similar to~\cite{8695705}, we use the central reference antenna to align the sequentially sampled signals.
The following assumes a simplified coplanar scenario with cyclostationary and frequency stable sources located in the far field of an isotropic $N$ element \ac{UCA}.
We model the simplified received signal vector $\mathbf{X_a} = [X_a^{m_1}\dots X_a^{m_N}]$ for each array element $m$ at location $[x_m, y_m]$ with the ideal \ac{UCA} steering vector as
\begin{multline}
    {X_a^m}(t) = \sum\limits_p a_p e^{j2\pi (\omega_p t + \varphi_p + \varphi_m)} \\
    \cdot e^{j2\pi ((x_m \cos(\Theta_p) + y_m \sin(\Theta_p))/\lambda_p)} + n_{\text{m}}(t),
    \label{eq:simSigAnt}
\end{multline}
where $p$ are the far-field incidence waves with amplitude $a_p$, phase $\varphi_p$, circular frequency and wavelength $\omega_p, \lambda_p$ and incidence angle $\Theta_p$.
$\varphi_m$ describes the random sampling phase.
The reference signal $X_r^m$ from location $x, y = 0$ is given by
\begin{equation}
    X_r^m(t) = \sum\limits_p a_p \cdot e^{j2\pi (\omega_p t + \varphi_p + \varphi_m)} + n_{\text{r}}(t).
\label{eq:simSigRef}
\end{equation}
The modeled noise $n_{\text{m}}(t)$, $n_{\text{r}}(t)$ in both cases is uncorrelated, zero-mean $\sigma\mathbf{I}$. 
The pseudo coherent signal $X_s^m(t)$ is
\begin{equation}
X_s^m(t) = X_a^m(t) \odot (X_r^m(t))^*.
\label{eq:coh}
\end{equation}

The resulting signal is constant and can be averaged over all samples $t$ with $\mathbf{T}$ being all sampled time steps for one antenna in order to speed up subsequent computations. 
The signal vector $\mathbf{R_X} \in \mathbb{C}^{N\times 1}$ gets composed as
\begin{align}
    \mathbf{R_X} &= [E\{\mathbf{X_s^{m_1}}\}, \dots, E\{\mathbf{X_s^{m_N}}\}]^T.
    \label{eq:avgOp}
\end{align}

\subsection{Array Compensation}
To enable spatial smoothing, the \ac{UCA} is transformed into a virtual \ac{ULA}, as described in~\cite{299559}.
Mutual coupling and array imperfections degrade performance when using the ideal transformation~\eqref{eq:virtTransform}.
The virtual array signal $\mathbf{\tilde{R}_X} \in \mathbb{C}^{N \times 1}$ is computed using a submatrix of the spatial DFT $\mathbf{F} \in \mathbb{C}^{(2h + 1) \times N}$.
Here, $h$ is the size of the smoothed sub-arrays, $r$ is the \ac{UCA} radius, $J_n$ is the Bessel function of order $n$, and $k$ is the wavenumber.

\begin{align}
    \mathbf{\tilde{R}_X} &= \mathbf{JF R_X}
    \label{eq:virtTransform}\\
    \mathbf{J}&=\operatorname{D}\left\{\frac{1}{\sqrt{N} j^n J_n(k r)}\right\} \quad n=-h, \cdots, 0, \cdots, h
\end{align}

Mutual coupling as well as array imperfections must be compensated for by calibrating the real steering vector.
This is done by probing the array with known incidence waves and optimizing the transformation and calibration matrix $\mathbf{B} \in \mathbb{C}^{N \times N}$.
$\mathbf{B}$ transforms the real, measured steering vectors $\mathbf{A} \in \mathbb{C}^{L \times N}$ with the number of probing angles $L$ into the virtual steering vectors.
\begin{align}
    \mathbf{A} &= \left[\mathbf{a}(\Theta_1), \dots,  \mathbf{a}(\Theta_L)\right]\\
    \mathbf{\tilde{A}} &= \left[\mathbf{\tilde{a}}(\Theta_1), \dots,  \mathbf{\tilde{a}}(\Theta_L)\right]\\
    &\min_{\mathbf{B}} ||\mathbf{\tilde{A}} - \mathbf{BA}||^2
    \label{eq:leastSQ}
\end{align}
For the virtual array steering vector, we use:
\begin{align}
    \mathbf{\tilde{a}}(\Theta) &= [e^{-jn\Theta}, \dots]^T \quad n=-\lfloor N/2\rfloor, \cdots, 0, \cdots, \lfloor (N-1)/2\rfloor.
    \label{eq:virtStruc}
\end{align}
A solution for the least-squares problem in~\eqref{eq:leastSQ} is given by
\begin{align}
    \mathbf{B}^H &= (\mathbf{AA}^H)^{-1}\mathbf{A\tilde{A}}^H.
\end{align}
Hence, we can compute the virtual array samples as 
\begin{align}
    \mathbf{\tilde{R}_x} = \mathbf{B}\mathbf{R_x}.
\end{align}
The virtual array is compensated for imperfections and mutual coupling and in a structure that is suitable for spatial smoothing.

\subsection{MUSIC evaluation}
The signal covariance matrix of the virtual array $\mathbf{C} \in \mathbb{C}^{N\times N}$ is calculated by
\begin{align}
    \mathbf{C} = \mathbf{\tilde{R}_x} \mathbf{\tilde{R}_x^H}
\end{align}

\subsubsection{Spatial Smoothing}
The forward smoothed covariance matrix $\mathbf{C_f} \in \mathbb{C}^{(N-h+1) \times (N-h+1)}$ is defined as follows:
\[
\mathbf{C_f} = \frac{1}{h} \sum_{i=0}^{h-1} \mathbf{C}[i:i+N-h, i:i+N-h],
\]
where  $\mathbf{C}[i:i+N-h, i:i+N-h]$  is the $(N-h+1) \times (N-h+1) $ submatrix of  $\mathbf{C}$ , at the $(i+1)$-th row and column.

Optionally, a backwards smoothing pass $\mathbf{C}_{\text{fb}}$ can be applied
\begin{align}
    \mathbf{C}_{\text{fb}} =\frac{1}{2}\left( \mathbf{C_f} + \text{flip}(\mathbf{C_f})^* \right).
\end{align}
where $\text{flip}(\mathbf{C_f})$ reverses rows and columns.

\subsubsection{MUSIC Spectrum}
Eigenvectors $V$ and Eigenvalues $W$ are computed by Singular Value Decomposition
\begin{align}
    \mathbf{W, V} = \text{svd}(\mathbf{C_{fb}}).
\end{align}
With $\mathbf{V_n} = \mathbf{V}[:, :N-h+1 - n_{\text{exp}}]$ the $n_{\text{exp}}$ 
eigenvectors associated with the smallest eigenvalues are taken as the noise subspace.

Finally, the MUSIC spectrum $\hat{P}_{\text{MUSIC}}(\Theta_t)$ is computed for all probing angles $\Theta_t$ using the ideal steering vector~\eqref{eq:virtStruc}.
\begin{align}
    \hat{P}_{\text{MUSIC}}(\Theta_t) &= \left| \frac{1}{\mathbf{\tilde{a}}(\Theta_t)^H \cdot \mathbf{V_n} \cdot \mathbf{V_n}^H \cdot \mathbf{\tilde{a}}(\Theta_t)} \right|
\end{align}

\section{Experiments}
\begin{figure}
    \centering
    \includegraphics[width=.8\linewidth]{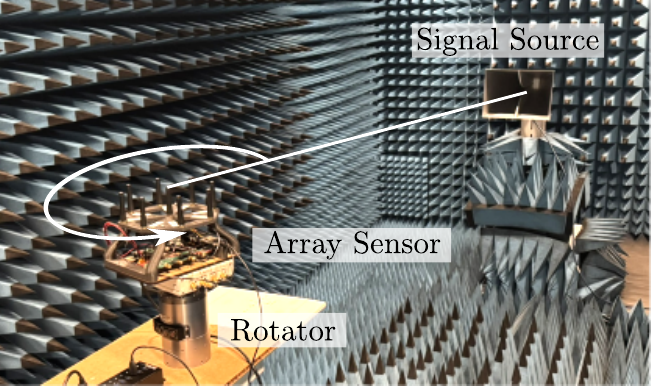}
    \caption{Measurement setup in the anechoic chamber. Measurements taken in \SI{10}{\degree} increments with the sensor mounted on a precision rotator.}
    \label{fig:measSetup}
\end{figure}

Quantitative experiments are performed in an anechoic chamber for reproducibility.
The sensor is mounted on a rotator at the bore sight of the signal source, as shown in Fig. \ref{fig:measSetup}.
Qualitative results in real-world multipath scenarios demonstrate the setup's robustness.
Calibration and testing data are collected separately to avoid compensating for slowly varying effects, such as phase changes from moving coaxial cables.
For all experiments, the calibration matrix $\mathbf{B}$ is generated using data spaced at $\Delta \Theta_p = 20^\circ$.

\subsection{Accuracy and \ac{SLL}}
For accuracy evaluation, measurements with $\Delta \Theta_p = 10^\circ$ are recorded.
Throughout the full rotation, a mean absolute angle error $E\{\epsilon\}$ of \SI{4.7}{\degree} with a standard deviation $\sigma_\epsilon$ of \SI{4.5}{\degree} is observed.

\begin{align}
        E\{\epsilon\} = \frac{1}{\lfloor 360/\Delta\Theta_p \rfloor} \sum_{i = 0}^{\lfloor 360/\Delta\Theta_p \rfloor} \left| \argmax_{\Theta_t}  (\hat{P}_\text{MUSIC}^{i\Delta\Theta_p}(\Theta_t)) - i\Delta\Theta_p \right|
\end{align}

\begin{figure}
    \centering
    \includegraphics[width=0.8\linewidth]{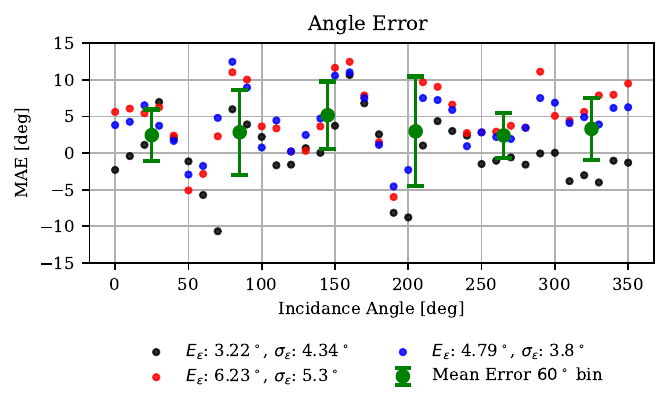}
    \caption{Angle errors for accuracy evaluation. Three datasets from anechoic chamber with mean and std.}
    \label{fig:accErrors}
    \vspace{-10pt}
\end{figure}

Fig~\ref{fig:accErrors} shows the angle errors of three datasets for each orientation.
Error bars indicate the mean error and std for bins of \SI{60}{\degree}.

\begin{figure}
    \centering
    \includegraphics[width=0.8\linewidth]{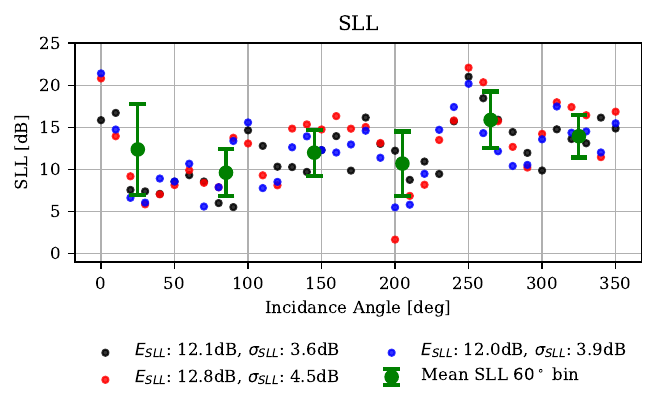}
    \caption{\ac{SLL} measurements from anechoic chamber.}
    \label{fig:snr}
    \vspace{-10pt}
\end{figure}

\ac{SLL} measurements are conducted with the same setup by comparing the main lobe power to the next highest peak.
Fig. \ref{fig:snr} shows the \ac{SLL} measurements for each measured angle together with \SI{60}{\degree} bucketed mean \ac{SLL} and std.
Over all measurements, the mean \ac{SLL} is \SI{12.3}{\dB} with \SI{4.1}{\dB} std.

\subsection{Real-World}
Empirical investigations are executed utilizing the ANYmal quadrupedal robot~\cite{Hutter2016ANYmalRobot} in indoor and dense urban outdoor settings. 
The sensor, mounted on the robot, tracks a handheld transmitter as illustrated in Fig. \ref{fig:anymal_aoa}. 
The degree of performance in estimation is significantly influenced by the intensity of multi-path reflections, rendering a reproducible, quantitative analysis impossible.
Although the estimated direction is mainly correct, its MAE precision is substantially compromised by sporadic deviations in direction and prominent side-lobes.
It has proven beneficial to low-pass filter the \acp{DoA} in order to smooth the occasional jumps.

\begin{figure}
    \centering
    \includegraphics[width=0.9\linewidth, clip, trim = 0 0 0 0mm]{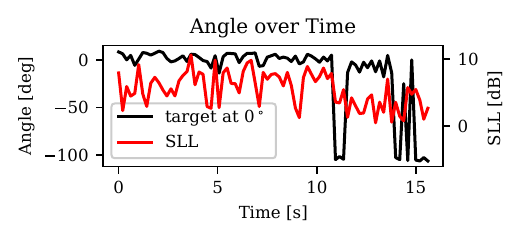}
    \caption{Outdoor measurement, target at $\Theta_p = 0^\circ$, increasing reflections.}
    \label{fig:multipath_degradation}
    \vspace{-10pt}
\end{figure}

Fig.~\ref{fig:multipath_degradation} shows the angle reading of a stationary target with increasing multi-path conditions.
At $t=11\text{s}$ MUSIC loses track, estimates a wrong angle and outputs a reduced \ac{SLL}.
At this time, multiple people are stepping into the scene, creating more reflections than the proposed setup can handle.
Real-world measurements generally show a lower \ac{SLL} of \SI{6}{\dB} at comparable conditions to the chamber measurements.

\section{Conclusion}
The proposed system demonstrates a cost-effective and practical approach to \ac{DoA} estimation using a software-defined radio with a switched \acp{UCA}. 
By leveraging pseudo-coherent sampling and an enhanced MUSIC algorithm with spatial smoothing, it achieves accurate results in both controlled and real-world environments.
However, as expected, its performance diminishes in highly reflective multipath scenarios, emphasizing the need for further improvements in such conditions. 
Overall, this work highlights the system's accessibility, robustness, and potential for deployment in a variety of RF localization tasks.

\newcommand{\IMSacktext}{

}

\IMSdisplayacksection{\IMSacktext}


\bibliographystyle{IEEEtran}

\bibliography{IEEEabrv,IEEEexample}

\end{document}